\documentstyle[12pt,aaspp4]{article}
\lefthead{Eskridge et al.}
\righthead{X-Ray Emission from M32}

\begin{document}

\title{X-Ray Emission from M32:  X-Ray Binaries or a ${\bf \mu}$AGN?}

\author{Paul B.~Eskridge, Raymond E.~White III, and David S.~Davis}
\affil{Department of Physics and Astronomy, University of Alabama, Tuscaloosa,
AL 35487}

\authoremail{paul@hera.astr.ua.edu, white@hera.astr.ua.edu, 
davis@hera.astr.ua.edu}

\begin{abstract}
We have analysed archival {\it ROSAT} PSPC data for M32 in order to study the
x-ray emission from this nearest elliptical galaxy.  We fit spectra from three 
long exposures with Raymond-Smith, thermal bremsstrahlung, and power-law 
models.  All models give excellent fits.  The thermal fits have kT$\approx$4 
keV, the Raymond-Smith iron abundance is $0.4^{+0.7}_{-0.3}$ Solar, the 
power-law fit has $\alpha$=1.6$\pm$0.1, and all fits have $N_H$ consistent with 
the Galactic column.  The source is centered on M32 to an accuracy of 9$''$, 
and unresolved at 27$''$ FWHM ($\sim$90 pc).  M32 is x-ray variable by a factor 
of 3--5 on timescales of a decade down to minutes, with evidence for a possible
period of $\sim$1.3 days.  

There are two plausible interpretations for these results:  1) Emission due to 
low-mass x-ray binaries; 2) Emission due to accretion onto a massive central 
black hole.  Both of these possibilities are supported by arguments based on 
previous studies of M32 and other old stellar systems; the {\it ROSAT} PSPC 
data do not allow us to unambiguously choose between them.  Observations with 
the {\it ROSAT} HRI and with {\it ASCA} are required to determine which of 
these two very different physical models is correct.
\end{abstract}


\section{Introduction}

M32 (NGC 221, PCG 2555) is not only the nearest (D$\approx$700 kpc, 
\markcite{cjfn}e.g.~Ciardullo et al.~1989) example of the low-luminosity end 
of the sequence of normal elliptical galaxies \markcite{korm1985}(Kormendy 
1985), but the nearest such elliptical of any luminosity.  This allows us to 
study it to much lower absolute luminosity and size limits than is possible for 
other more luminous ellipticals.  The Virgo cluster is $\sim$20 times more 
distant than M32 \markcite{p94}(e.g.~Pierce et al.~1994), thus observations at 
a given angular scale or flux limit probe $\sim$400 times deeper for M32 than 
for Virgo Es.  As a result, our understanding of the properties of M32 provides 
us with a cornerstone for the investigation of more luminous and more distant 
ellipticals.

X-ray emission from faint early-type galaxies ($L_B \la 10^{40}~{\rm 
erg~s^{-1}}$) typically has a hard ($\sim$5 keV) spectrum, with $L_X \propto 
L_B$ \markcite{kft92a}(Kim, Fabbiano \& Trinchieri 1992a; \markcite{efka1995} 
Eskridge, Fabbiano \& Kim 1995).  This emission is interpreted as arising from 
population II stellar binary sources (low mass x-ray binaries, or LMXRBs).  The 
{\it Einstein} data for M32 are consistent with this interpretation 
\markcite{fkt1992}(e.g.~Fabbiano, Kim \& Trinchieri 1992), however the {\it 
Einstein} observation of M32 was not particularly deep; other models for the 
x-ray emission are not excluded.  In particular, optical imaging and 
spectroscopy both indicate that M32 contains a central black hole of 
$M_{\bullet} \approx$ 1--3$\times 10^6M_{\odot}$ (van der Marel et al.~1994; 
Lauer et al.~1992).  The observed x-ray emission may be due to low-level 
accretion onto this black hole.

We have analysed archival {\it ROSAT} PSPC observations of M32 in order to 
study the nature of the x-ray emission with the best currently available data.  
In \S 2 we discuss the available PSPC data.  In \S 3 we investigate the 
spectral properties, extendedness, and time variability of the x-ray emission.  
We discuss the two possible physical interpretations of these data in \S 4, and 
conclude with a discussion of crucial future observations in \S 5.

\section{PSPC Observations}

There are many pointings that include M32 in the {\it ROSAT} Public Archive.  
Most are short ($\sim$2--3 ksec), but there are three long unobstructed 
pointings ($\ga$28 ksec), details of which are given in Table 1.  The total 
integration time from these pointings is $\sim$101 ksec.  We use these data for 
spectral analysis in \S 3.1 below.  We use two pointings to analyse the source 
extent in \S 3.2:  WP600363N00 is the pointing nearest to on-axis but is only a 
$\sim$2 ksec exposure; WP600068 (17$'$ off-axis) is the long pointing nearest 
to on-axis.  We use the short pointings to search for source variability in \S 
3.3 below.

\section{Analysis}

\subsection{Spectral Properties}

We fit the vignetting-corrected data from the three long pointings to spectral 
models using {\sl XSPEC v9.0}.  The source count-rate is $\sim$0.1${\rm 
s^{-1}}$.  Source spectra were extracted from circular regions centered on M32,
with radii of $2'\llap.3$--$5'\llap.2$ (or 3.8--8.7$R_e$ for $R_e = 0'\llap.6$, 
\markcite{rc3}de Vaucouleurs et al.~1991) depending on the off-axis distance of
M32 in each pointing.  We adopted a maximum master-veto rate of 170.  
Background regions were chosen from adjacent source-free areas.  Errors were 
calculated using Poisson statistics, with systematic errors of 1\% included in
quadrature.  We fit three kinds of spectral models to the data:  Raymond-Smith 
(RS) thermal models, thermal bremsstrahlung (TB), and power-law (PL) models.  
These models incorporate an absorption component with \markcite{mcs}Morrison \& 
McCammon (1983) cross-sections.  As M32 exhibits significant variability (see 
\S 3.3 below), the normalizations for each pointing were allowed to vary 
independently.

All three spectral models provide formally excellent fits to the data, with the 
RS model being the best by a small margin (see Table 2).  The best-fit RS model
has kT$\approx$3.9 keV, and an abundance of $\sim$0.4 Solar (with large error 
bounds).  The best-fit TB model has kT$\approx$4.3 keV.  The best-fit PL model 
has a photon index of $\alpha \approx$1.6.  The best-fit RS model is shown with 
the three data sets (with different normalizations) in Figure 1a,b.  The 
best-fits to the data for the three spectral models are essentially 
indistinguishable.  The fit $N_H$ is consistent with the observed Galactic 
column ($N_H\approx 6.6\times10^{-20}$ cm$^{-2}$ \markcite{hi92}Stark et 
al.~1992) for all models.  The TB model has a higher $\chi^2$ than the RS 
model:  $\Delta\chi^2=+5.8$ for $\Delta\nu=+1$.  The $F$-test (which tests if 
the reduction in $\chi^2$ due to an additional fitted parameter is significant) 
suggests that the line-producing RS model is preferred to the zero-abundance TB 
model at the $\sim$98\% level.  The fluxes and luminosities implied by these 
spectral models are given in Table 3 for both {\it ROSAT} and {\it Einstein} 
energy bands.  The fluxes from the three pointings differ by $\sim$10\%, an 
amount much larger than the change in flux between different models for a given 
pointing.  We believe this is due to the intrinsic variability of the source 
(discussed in \S 3.3, below).

We searched for evidence of emission from gas at or near the kinetic 
temperature of the stars (40$\leq$kT$\leq$200 for a central velocity dispersion 
of 77 km s$^{-1}$, \markcite{McE1995}McElroy 1995), but found that the flux 
attributable to any such cool component is $\la$1\% of the total.

\subsection{Extendedness}

The on-axis exposures are all $\la$3 ksec, and thus do not provide good 
counting statistics for evaluating the extendedness of M32.  Therefore, we use 
both WP600363N00, a $\sim$2 ksec exposure with M32 3$'$ off-axis, and WP600068, 
a 30.6 ksec exposure with M32 17$'$ off-axis.  The central region of WP600068 
is displayed in Figure 2a.  We used {\sl PROS} to produce this image from the
initial data file, and {\sl ftools v3.3} to generate the exposure map.  Figure 
2b is a contour-plot of M32,  and shows an extension to the NE due to a second, 
partially resolved faint source (with a count-rate $\la$4\% that of the main 
source).  The nature of this second source is unclear.  We fit the radial 
profile of M32 in WP600068 (excluding $0^{\circ} \leq PA \leq 90^{\circ}$ in 
order to avoid the second source) with a PSF appropriate for a 17$'$ off-axis 
PSPC observation \markcite{has}(Hasinger et al.~1995).  This PSF has a 
FWHM$\approx$46${''}\llap.3$ (1.3$R_e$ or $\sim$160 pc).  The profile is 
well-fit by the PSF (see Figure 3).  A similar analysis of WP600363N00 
indicates that M32 is unresolved at 27$''$ FWHM ($\sim$0.75 $R_e$, or $\sim$90 
pc), although with much poorer counting statistics.  We measured the position 
of the M32 x-ray source in the image WP600342.  Using the point-sources 5C 
03.076 and M31 40 1650\footnote{This star is just west of NGC 206.  Its optical 
and x-ray luminosities are consistent with it being a high-mass x-ray binary in 
the disk of M31.} to transform the astrometry, we derive an x-ray position for 
M32 of $00^h42^m41^s\llap.7$, $+40^{\circ}51'45{''}\llap.1$ (J2000).  This 
differs from the RC3 optical position by $\sim$9$''$, comparable to the PSPC 
positional accuracy.

\subsection{Time Variability}

For a 4 keV TB model the average flux from the PSPC implies a flux in the {\it 
Einstein} band that is $\sim$3 times that observed in 1980 (the other spectral 
models give the same result -- see Table 3).  Thus M32 is x-ray variable on the 
timescale of a decade.  In Figure 4a, we show the count-rate time series for 
all PSPC pointings with M32 $\leq$40$'$ off-axis.  These data are sparsely 
sampled, but show clear variability with an amplitude of $\sim$1 dex on 
timescales of months to a year.  Fig.~4b shows the best-sampled part of the 
time series.  We fit these data with a phase dispersion minimization algorithm, 
and find evidence for a period of 1.27 days.  A second moderately well-sampled 
part of the time series, around day 540, shows evidence for a similar period 
(1.37 days), but this is based on only twelve data points.  We attempted to 
phase these two samples together, searching for a period in the range 1.1--1.5 
days.  None was found.  Thus the available data show no sign of a coherant 
binary period over a gap of $\sim$200 days, arguing that the evidence for 
periodicity in the short samples may be spurious.  We also studied the 
count-rate data for the pointing WP600068 to search for shorter timescale 
variability.  We note that although M32 is near the PSPC ring structure in this 
exposure, it is {\it not} close enough to be significantly effected by the 
spacecraft wobble.  While both the Kolmogorov-Smirnov and the Cramer-von Mises 
tests argue that the source is variable (at the $>$99\% and 95\% levels, 
repectively), we find no convincing period aside from the orbital period 
($\sim$90 min).

\section{Discussion}

The x-ray data for M32 suggest two possible interpretations:  1) The x-rays are 
due to the integrated emission from a population of LMXRBs.  2) The x-rays are 
due to emission from accretion onto a massive central black hole.  Below, we 
review the available evidence for and against each of these interpretations.

\subsection{The Case for LMXRBs}

The $L_X$ for M32 corresponds to the Eddington luminosity ($L_E$) for a 
1--3$M_{\odot}$ object, and is in keeping with the $L_X$--$L_B$ relationship 
for x-ray faint early-type galaxies \markcite{efka1995}(Eskridge et al.~1995).  
The standard interpretation for x-ray emission from such galaxies is that it is 
due to LMXRBs \markcite{kft92b}(e.g.~Kim, Fabbiano \& Trinchieri 1992b).  There 
is an extensive literature on the stellar populations of M32 
\markcite{buzz95}(e.g.~Buzzoni 1995, and references therein).  Although many 
studies indicate a significant intermediate age population exists 
\markcite{buzz95}(references in Buzzoni 1995), the dominant population in terms 
of mass appears to be old.  Thus one would expect a population of LMXRBs.  The 
best-fit thermal model to the {\it ROSAT} spectrum has kT$\sim$4 keV, 
consistent with results for Galactic LMXRBs \markcite{xrb95}(e.g.~White, Nagase 
\& Parmar 1995).  The RS abundance is $0.1 \leq Z_{\odot} \leq 1.2$.  Estimates 
from integrated nuclear spectroscopy are in the range $-0.5 \leq [Fe/H] \leq 
-0.1$ (0.3 --- 0.8 $Z_{\odot}$) \markcite{hccj94}(e.g.~Hardy et al.~1994 and 
references therein).  The x-ray result thus agrees with the optical results.  
We noted above that M32 is x-ray variable on timescales from a decade down to a 
day.  Galactic LMXRBs are x-ray variable on $\sim$day timescales due to the 
binary periods, and exhibit x-ray flaring and quasi-periodic behaviour on 
longer timescales \markcite{xrb95}(e.g.~White et al.~1995).  The number of 
LMXRBs needed to account for the observed x-ray flux from M32 is small, only 
1--5 objects, thus the observed global variability is consistent with the LMXRB 
interpretation. 

\subsection{The Case for a ${\bf \mu}$AGN}

The {\it ROSAT} data are as well fit by an $\alpha$=1.6$\pm$0.1 PL model as by 
a thermal model.  An $\alpha$=1.7 PL is the canonical x-ray spectrum for AGN 
\markcite{mdp93}(e.g.~Mushotzky, Done \& Pounds 1993).  AGN are well known to 
be x-ray variable \markcite{mdp93}(e.g.~Mushotzky et al.~1993) on the 
timescales and with the amplitudes observed.  There is no evidence that the 
main source is extended at the level of 0.75$R_e$.  We are thus forced to take 
seriously the possibility that the x-ray emission from M32 is due to a 
$\mu$AGN.  

The possible existence of a massive central black hole in M32 has been 
discussed for the last dozen years \markcite{vdm94}(e.g.~van der Marel et 
al.~1994 and references therein).  The relevant data are the nuclear rotation 
and velocity dispersion profiles, and the nuclear luminosity profile.  
\markcite{vdm94}Van der Marel et al.~(1994) analyse high spatial and spectral 
resolution, high S/N nuclear spectra of M32.  They conclude that their data can 
be explained by the presence of a central black hole of $M_{\bullet} \approx 
1.8 \times 10^6 M_{\odot}$.  Imaging of the central luminosity cusp at a 
resolution of $0{''}\llap.04$ by \markcite{lau92}Lauer et al.~(1992) supports 
the existence of a central black hole of $M_{\bullet} \approx 3 \times 10^6 
M_{\odot}$.  These studies also strongly constrain non-black hole models, in 
that the relaxation and stellar collision timescales for extended mass 
distributions that can produce their results are $\sim$10$^{-2}$ of a Hubble 
time.  

The $L_E$ for a black hole of $M_{\bullet} \approx 3 \times 10^6 M_{\odot}$ is 
$\sim$4$\times 10^{44}~{\rm erg~s^{-1}}$, six dex higher than the observed 
$L_X$.  The low $L_X$ can be attributed to a lack of fuel for the central 
engine:  $M_{HI} < 5 \times 10^5M_{\odot}$ \markcite{rhb91}(Roberts et 
al.~1991); $L_{FIR} < 3 \times 10^{36}~{\rm erg~s^{-1}}$ 
\markcite{knap89}(Knapp et al.~1989).  This is also reflected in the lack of 
any other observational signals of an AGN:  $P_{6cm} \leq 1$ mJy 
\markcite{rhb91}(Roberts et al.~1991); the limit to the equivalent width of 
H$\alpha$ is $<$6\AA~in a 3$''$ aperture \markcite{kk83}(Kennicutt \& Kent 
1983), or $\leq 1.8$\AA~in a 1$''$ aperture \markcite{keel83}(Keel 1983).  This 
implies $f_{H\alpha} \la 10^{-15}~{\rm erg~cm^{-2}~s^{-1}}$ (W.~Keel, private 
communication).  The $f_{H\alpha}/f_X$ ratio is moderately lower than in 
typical AGN \markcite{bkm}(e.g.~Blumenthal, Keel \& Miller 1982; 
\markcite{mea}Mulchaey et al.~1994), but not strikingly so.  If we take the 
observed $L_X \approx 10^{38}~{\rm erg~s^{-1}}$ as the bolometric luminosity of 
the central engine, we can use the well known relationship, 
$$\dot{M} \approx {{2 L}\over c^2}, \eqno{(1)}$$
to determine a mass accretion rate of 1--2$\times 10^{-9}M_{\odot}~{\rm 
yr^{-1}}$.  Such a fueling rate clearly offers no conflict with the low 
observational limits on AGN activity or ISM content in other wavebands.  
Finally, we note that if the x-rays {\it are} due to LMXRBs, then the putative 
central black hole is being fueled at a much lower rate than 
$10^{-9}M_{\odot}~{\rm yr^{-1}}$.

\section{Conclusion}

We have presented arguments that the x-ray emission from M32 can be due to {\it 
either} a small population of LMXRBs, {\it or} accretion onto the massive 
central black hole.  Two observations that could resolve this ambiguity are 
{\it ROSAT} HRI imaging and {\it ASCA} spectroscopy.  A deep on-axis HRI image 
would determine the extent and position of the M32 x-ray source much more 
precisely than do the PSPC data.  It would also resolve the weak source to the 
NE.  The HRI on-axis PSF is $\sim$6$''$ FWHM ($\sim$0.17$R_e$) giving a spatial 
resolution of $\sim$20 pc.  Emission from an AGN would thus be unresolved with 
the HRI, and would come from the center of M32.  Emission from a collection of 
LMXRBs would not have to be coincident with the center of M32, and could be 
extended at HRI resolution.  A deep {\it ASCA} spectrum will have the spectral 
resolution to clearly distinguish between the thermal and power-law spectral 
models in an integration time of $\sim$20 ksec.  Given the importance and 
proximity of M32, the opportunity to resolve the current puzzle should not be 
missed.

\acknowledgments

We thank Bill Keel for access to his unpublished nuclear spectrum of M32, and 
for useful discussions about AGN.  We also thank Jeff McClintock and Andy 
Silber for advice on the properties of LMXRBs.  This research has made use of 
the NASA/IPAC Extragalactic Database (NED) which is operated by the Jet 
Propulsion Laboratory, Caltech, under contract with the National Aeronautics 
and Space Administration.  This research was supported by the EPSCoR program 
under Grant No.~EHR-9108761, and by the National Aeronautics and Space 
Administration under ROSAT Grants No.~NAG 5-1973 and NAG 5-1718.

\newpage 

{
\baselineskip12pt
\tolerance=500
 
\def\tabrule{\noalign{\hrule}}
\def\pz{\phantom{0}}
\def\pb{\phantom{-}}
\def\pd{\phantom{.}}
\ 
 
\centerline{Table 1 -- Long {\it ROSAT} PSPC Pointings Including M32} 
\vskip0.3cm
 
\newbox\tablebox
\setbox\tablebox = \vbox {
 
\halign{\pz\pz#\hfil&&\hfil\pz#\pz\hfil&\hfil\pz#\pz\hfil&\hfil
\pz#\pz\hfil&\hfil\pz#\pz\hfil\cr
\tabrule
\noalign{\vskip0.1cm}
\tabrule
\noalign{\vskip0.1cm}
 
Seq.~ID & Exp.~time & RA(J2000) & Dec(J2000) & Offset Angle \cr
& sec & $\pz\pz^h\pz\pz^m\pz\pz^s$ & $\pz\pz^{\circ}\pz\pz'\pz\pz''$& \cr
\noalign{\vskip0.1cm}
\tabrule
\noalign{\vskip0.2cm}
WP600068 & $\pz$30594 & 00 42 28 & 41 08 24 & 17$'$ \cr
WP600067 & $\pz$27970 & 00 43 55 & 41 30 36 & 41$'$ \cr
WP600079 & $\pz$42164 & 00 39 36 & 40 24 00 & 45$'$ \cr
\noalign{\vskip0.2cm}
\tabrule
\noalign{\vskip0.2cm}
Total integration & 100728 & & & \cr
\noalign{\vskip0.2cm}
\tabrule
}
}
\centerline{ \box\tablebox}
}

{
\baselineskip12pt
\tolerance=500
 
\def\tabrule{\noalign{\hrule}}
\def\pz{\phantom{0}}
\def\pb{\phantom{-}}
\def\pd{\phantom{.}}
\ 
 
\centerline{Table 2 - Model Fitting Results}
\vskip0.3cm
 
\newbox\tablebox
\setbox\tablebox = \vbox {
 
\halign{#&\hfil\pz#\hfil&\hfil\pz#\hfil&\hfil\pz#\hfil
&\hfil\pz#\hfil&\hfil#\hfil&\hfil\pz#\hfil&\hfil#
\hfil&\hfil#\hfil&\hfil\pz#\hfil\cr
\tabrule
\noalign{\vskip0.1cm}
\tabrule
\noalign{\vskip0.1cm}
 
Model & kT & range$^1$ & $Z$ & range$^1$ & $\alpha$ & range$^1$ & $N_H$ &
range$^1$ & $\chi^2/\nu$ \cr
\ & keV & & $Z_{\odot}$=1 & & & & $10^{-20}~{\rm cm^{-2}}$ & & \cr
\noalign{\vskip0.1cm}
\tabrule
\noalign{\vskip0.2cm}
R-S & 3.94 & 2.88--5.67 & 0.44 & 0.13--1.16 & & & 7.18 & 6.31--$\pz$8.34 &
$\pz$89.5/104=0.86 \cr
Bremm. & 4.25 & 3.07--6.01 & & & & & 7.71 & 6.81--$\pz$9.05 & 
$\pz$95.3/105=0.91 \cr
Power-law & & & & & 1.57 & 1.46--1.71 & 8.45 & 7.29--10.23 & 102.0/105=0.97 \cr
\noalign{\vskip0.1cm}
\tabrule
}
}
\centerline{ \box\tablebox}

\vskip15pt

1.  The ranges given are the 90\% confidence-limits.
} 

{
\baselineskip12pt
\tolerance=500
 
\def\tabrule{\noalign{\hrule}}
\def\pz{\phantom{0}}
\def\pb{\phantom{-}}
\def\pd{\phantom{.}}
\ 
 
\centerline{Table 3 - Derived Fluxes and Luminosities}
\vskip0.3cm
 
\newbox\tablebox
\setbox\tablebox = \vbox {
 
\halign{\pz\pz#\pz&\hfil\pz#\pz\hfil&\hfil\pz#\pz\hfil&\hfil\pz#\pz\hfil&\hfil
\pz#\pz\hfil&\hfil\pz#\pz\hfil\cr
\tabrule
\noalign{\vskip0.1cm}
\tabrule
\noalign{\vskip0.1cm}
 
Model & Pointing & Flux$^1$ & Luminosity$^2$ & Flux$^1$ & Luminosity$^2$ \cr
 & & 0.2--2.0 keV & 0.2--2.0 keV & 0.5--4.0 keV & 0.5--4.0 keV \cr
\noalign{\vskip0.1cm}
\tabrule
\noalign{\vskip0.2cm}
Raymond-Smith & WP600079 & 1.92 & 1.13 & 3.24 & 1.90 \cr
 & WP600067 & 1.85 & 1.08 & 3.13 & 1.84 \cr
 & WP600068 & 1.61 & 0.94 & 2.72 & 1.60 \cr
\noalign{\vskip0.2cm}
Bremmstrahlung & WP600079 & 1.95 & 1.14 & 3.36 & 1.97 \cr
 & WP600067 & 1.88 & 1.10 & 3.24 & 1.90 \cr
 & WP600068 & 1.64 & 0.96 & 2.83 & 1.66 \cr
\noalign{\vskip0.2cm}
Power-Law & WP600067 & 1.97 & 1.16 & 3.67 & 2.15 \cr
 & WP600067 & 1.90 & 1.11 & 3.53 & 2.07 \cr
 & WP600068 & 1.65 & 0.97 & 3.08 & 1.81 \cr
\noalign{\vskip0.2cm}
\tabrule
}
}
\centerline{ \box\tablebox}
 
\vskip 15pt
 
1.  Flux in units of $10^{-12}~{\rm erg~s^{-1}~cm^{-2}}$

2.  Luminosity in units of $10^{38}~{\rm erg~s^{-1}}$ for an assumed distance 
of 700 kpc.
}

\newpage 

\figcaption{The {\it ROSAT} PSPC spectra for M32 from three long pointings.  a)
Data from the three pointings with the best-fitting RS model.  The data from
WP600068 have the highest amplitude, those from WP600067 have the lowest.  b)
$\chi^2$ as a function of energy for the RS fit.}

\figcaption{a) The central 50$' \times 50'$ of {\it ROSAT} image WP600068.  
North is up, East to the left.  The image center is at $\alpha = 
00^h42^m28^s\llap.7$, $\delta = +41^{\circ}08'24''$ (J2000).  The complex 
emission at ($-150''$,$490''$) is the central area of M31.  M32 is $\sim$17$'$ 
off-axis to the south.  b) A contour-plot of the $6'\llap.25 \times 6'\llap.25$ 
region centered on M32.  The low contour is at 6 counts per pixel (a 
significance level of 2.3$\sigma$), with subsequent contours at factor of 2 
intervals .  Note the obvious extension to the NE caused by a second, much 
fainter source.}

\figcaption{The radial profile of M32 from {\it ROSAT} image WP600068 with the
range $0^{\circ} \leq PA \leq 90^{\circ}$ excluded.  The solid line is the PSF 
for a source 17$'$ off-axis. } 

\figcaption{The PSPC light-curve for M32. a) All data from pointings 
$\leq$40$'$ off-axis.  b) The moderately well sampled part of the light curve
around UTC 388.  These data show evidence for a periodicity of $\sim$1.27 days.
}

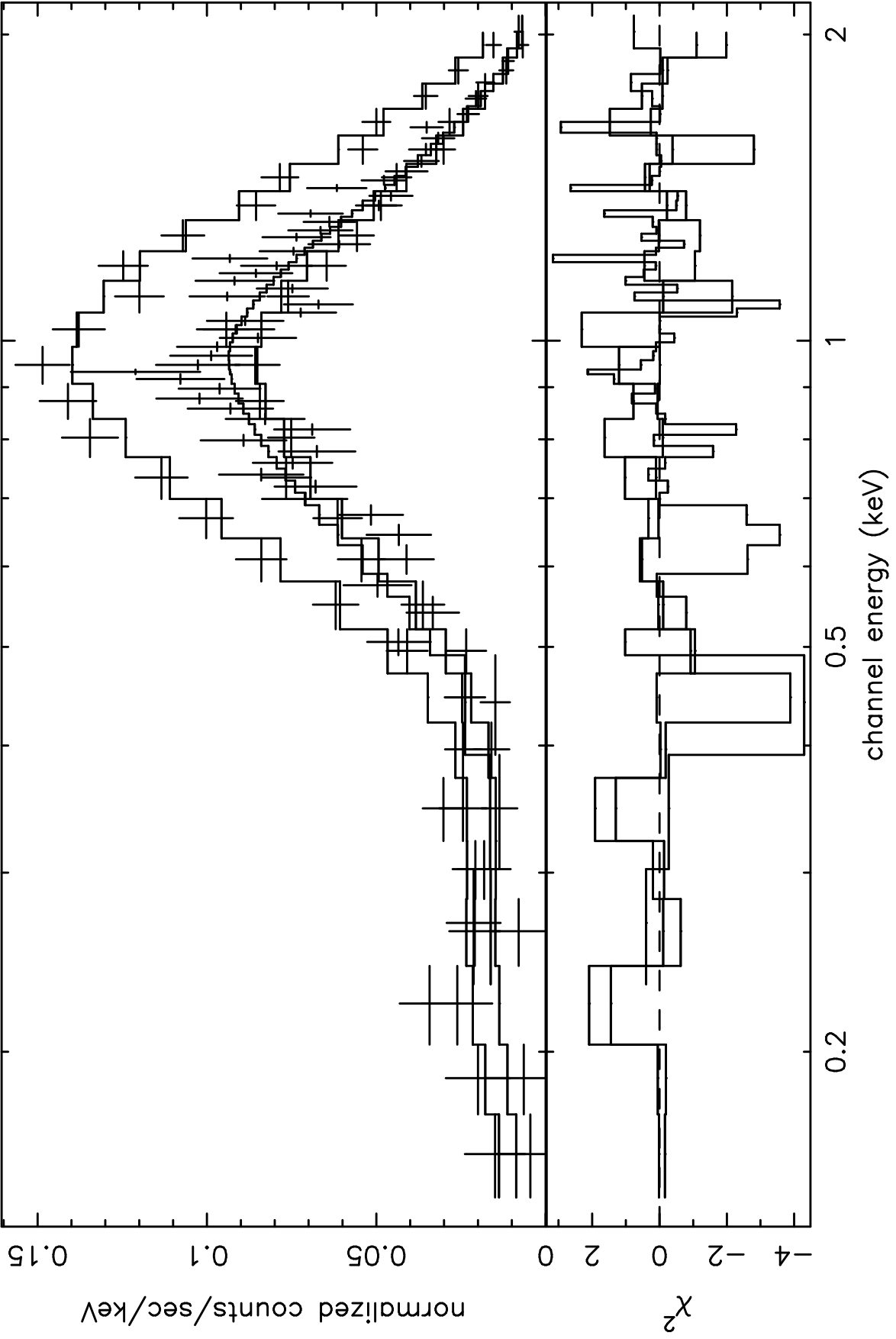
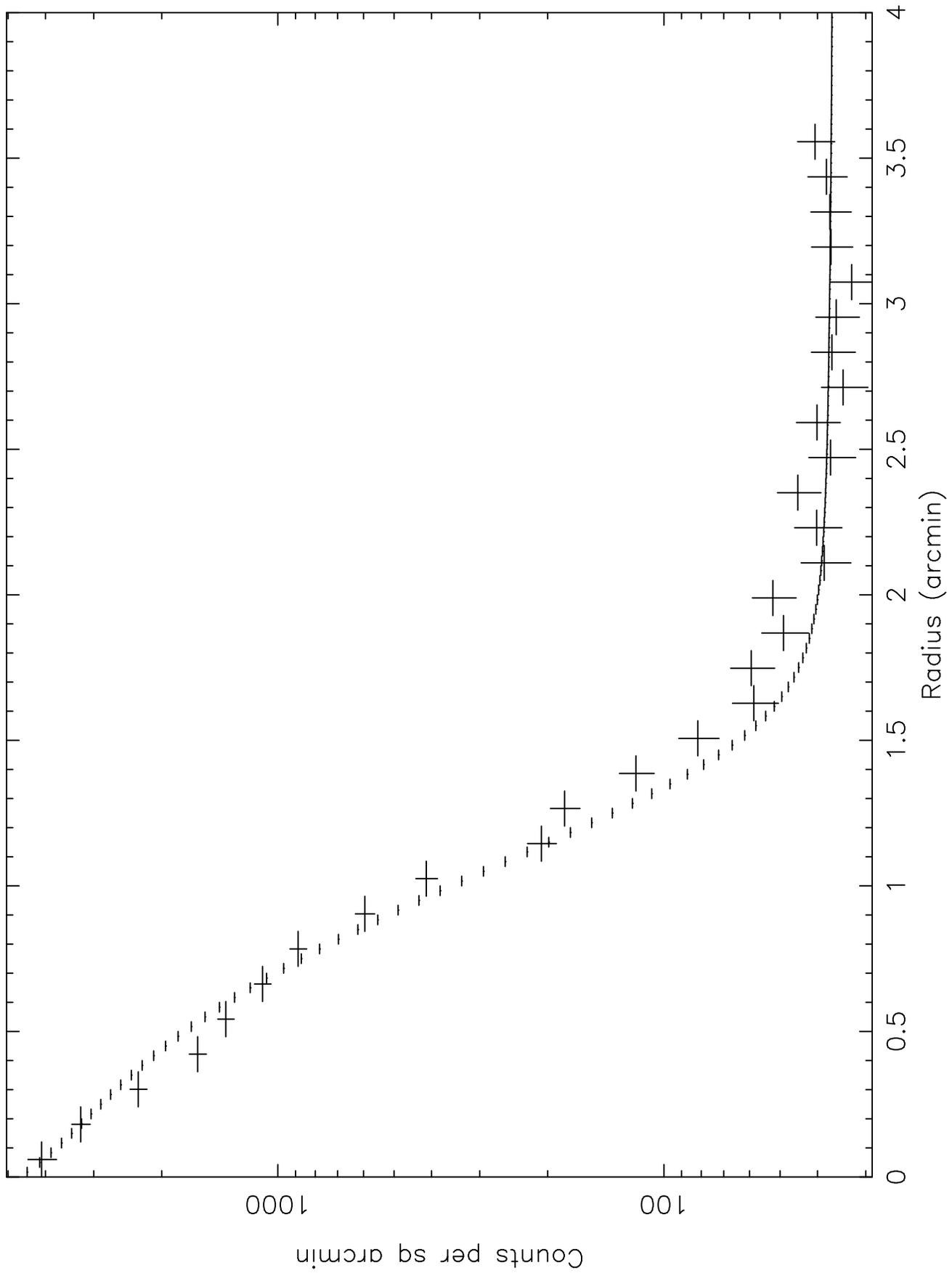
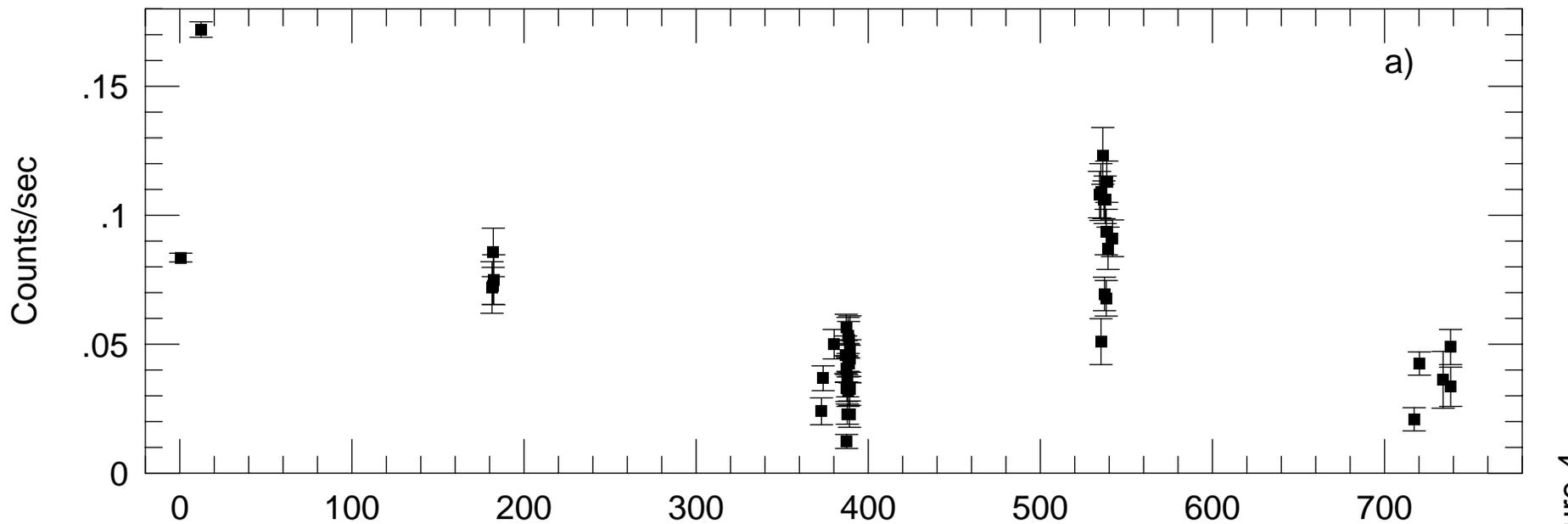
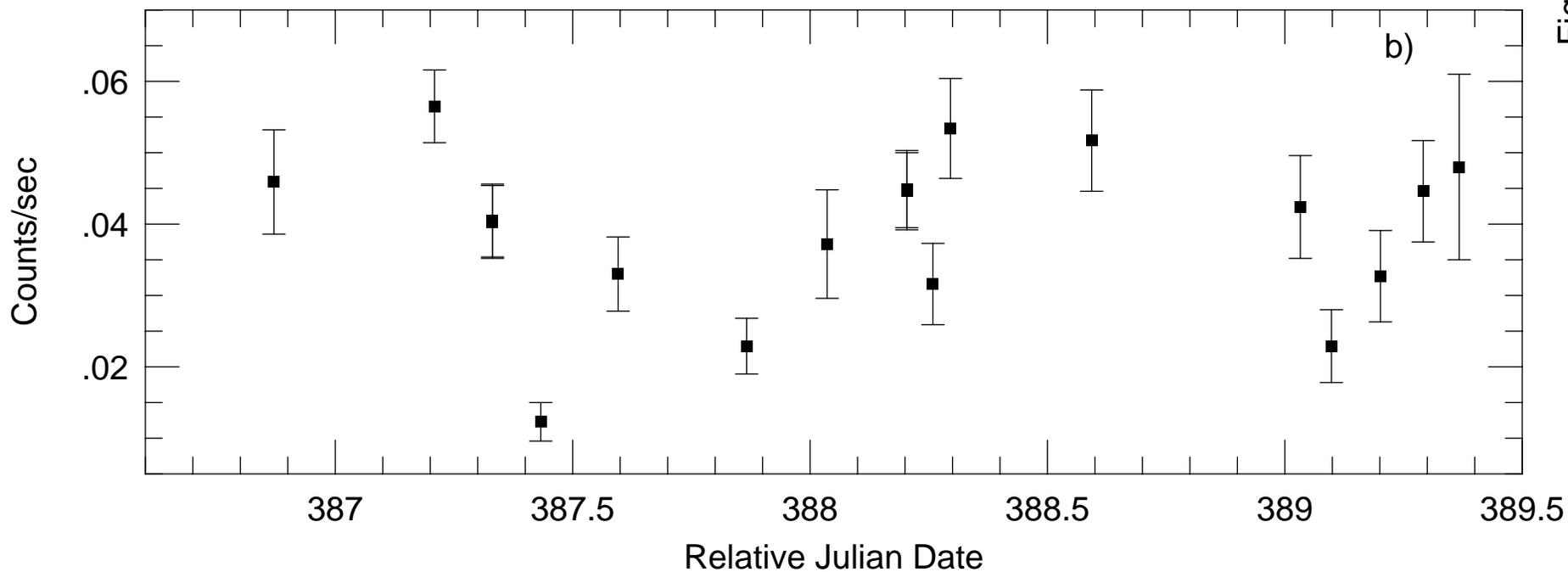
\end{document}